\documentclass[conference,compsoc]{IEEEtran}
\ifCLASSOPTIONcompsoc
\usepackage[nocompress]{cite}
\else
\usepackage{cite}
\fi \ifCLASSINFOpdf \else \fi

\usepackage{graphicx}

\begin{document}
\title{Energy-Efficient Strip Monitoring by Identical Devices
Directed to One Side Along the Strip and Having a Coverage Area in
the Form of a Sector}
\author{\IEEEauthorblockN{Adil Erzin}
\IEEEauthorblockA{Sobolev Institute of Mathematics\\Novosibirsk
State University\\Novosibirsk, Russia\\} e-mail:
adilerzin@math.nsc.ru}

\maketitle

\IEEEpeerreviewmaketitle

\section{Introduction}
Wireless sensor networks (WSN) are often used in security systems
to observe (to monitor) the territory
\cite{ZalErAsChoo09,AsErZal09,AsEr09}. If the object of
observation is a road, a pipeline or a corridor, it can be assumed
that the strip is monitored
\cite{AsEr09,ErAs11,AsEr13,AsEr13Vych}. The sensor has a coverage
area of various shapes within which it collects data. If the
sensor is equipped with a video camera, then depending on its
location, the coverage area on the surface may be a disk, an
ellipse or a sector. In the WSN, however, as well as in wired
sensor networks, a scarce resource is the energy of the sensors,
which is consumed proportionally to the area covered by them
\cite{ZalErAsChoo09,AsEr09,CarleSim04,FanJin10}. If the sensor
coverage areas overlap, this means that extra energy is wasted.
Thus, the problem of energy-efficient monitoring (sensing) is
reduced to the problem of computational geometry consisting in the
search for the least density coverage
\cite{Tot95,ZalErAsChoo09,DesSh13}.

Usually in the literature \emph{regular} covers are considered
\cite{ZalErAsChoo09,AsEr09,Kershner1939,Tot95}. In a regular cover
from class $COV_k(p,q)$, the region is divided into the equal
$k$-gons, which are called \emph{tiles}, and all the tiles are
covered equally with $p$ figures of $q$ different types
\cite{AsErZal09}. To analyze the density of a regular cover, it
suffices to consider the covering of one tile.

In the covers of a flat area, various figures can be used. Until
recently, circles (disks) were used as figures in the covers
\cite{FanJin10,Kershner1939,Tot95,CarleSim04,CardeiWuLu06,GuvYav11,IsKim14}.
Recently, coverings using ellipses \cite{AsEr13Vych,ErAs13} and
sectors
\cite{GuvYav11,Ai06,MaLiu07,HanCa08,DesSh13,LiangLo15,ErzinShab15,ErzinOsot16,Erzin16,Erzin16,MohShR14}
have also been considered.

In most publications, the area to be covered is the whole plane
\cite{Tot95,AsErZal09,ErAs13,FanJin10,Kershner1939,CarleSim04}.
The models of covering bounded regions is more complicated.
Effective attempts were made in the construction of strip
coverings with circles \cite{AsEr09,ErAs11,AsEr12}, ellipses
\cite{AsEr13} and sectors
\cite{GuvYav11,Ai06,MaLiu07,HanCa08,LiangLo15,ErzinShab15,ErzinOsot16}.

It was possible to prove the optimality of the covers only in some
classes \cite{Kershner1939,AsErZal09}. For most covers, accurate
estimates of accuracy are found
\cite{Tot95,AsEr09,ErAs11,AsEr12,ErAs13,ErzinShab15,ErzinOsot16}.

In \cite{ErzinShab15} the problem of constructing a least density
cover of a strip with identical sectors is studied. Several
coverage models are proposed and their comparative analysis is
performed which, in particular, allows one to obtain an upper
bound for the minimum density of a strip coverage with equal
sectors. In \cite{Erzin16} the problem of constructing a
cost-effective cover of a strip with identical sectors is
considered, and several efficient coverage models are investigated
and their comparative analysis is performed.

In the ordinary cover it is sufficient to cover each point of the
strip, regardless of the orientation of the sensors covering the
area. Another situation arises when it is necessary to look at the
objects from a certain side. For example, in video surveillance
systems on the roads, or to identify people walking along the
corridor (to passport control or boarding a plane), it is
necessary to see their faces. This means that the video cameras
must be directed towards the faces of walking people.

Many modern sensors can adjust their coverage area, and the
density can be reduced by choosing the optimal values for the
parameters of the figures involved in the covering
\cite{Tot95,ZalErAsChoo09,CardeiWuLu06,MohShR14}.

In this paper, we propose several regular covers with identical
sectors that observe the strip in the same direction, and we
carried out their analysis, which allows choosing the best
coverage model and the best parameters of the sector.

The rest of the paper is organized as follows. The mathematical
formulation of the problem is given in Section 2. In Section 3, a
coverage model S1 is considered. A coverage model S2 is considered
in Section 4. A coverage model S3 is presented in Section 5, and
the paper is concluded in Section 6.

\section{Problem Formulation}
We are given a strip whose width, without loss of generality, is
set equal to 1, and we need to observe the objects that are being
drifted along the strip, for example, from left to right. Let the
sensor's coverage area be a sector $(R,\alpha)$, where $R$ is the
radius and $\alpha$ is the angle, in the vertex of which the
sensor is located. The sensor can be placed anywhere in the strip
and can be oriented in any direction.

The point of the strip is \emph{covered} if it belongs to at least
one sector whose vertex is not to the left of the point.

A \emph{cover} is such placement of sectors, and setting their
orientations, that each point of the strip is covered.\\

\textbf{\emph{The problem is to find the cover of the strip of minimum
density.}}\\

Let us call the coverage \emph{model} the mutual arrangement of
sectors. If we set the parameters of the sector in the coverage
model, we get a specific cover. We consider three coverage models
S1, S2 and S3 and find the optimal parameters of the sector for
each model, thereby determining the concrete covers S1, S2 and S3.

Only three coverage models are considered in this paper, because
they seem to be the most effective. It can be shown that other
covers can be reduced to the covers under consideration without
increasing the density.

\section{Coverage Model S1}
In the model S1, the sector angle does not exceed $\pi/3$ and the
inequality $R\sin\alpha\geq 1$ holds. One side of the sector lies
on the border of the strip, the other side crosses the opposite
boundary of the strip, and the sector is oriented to the left. The
next sector on the right is symmetric about the center line to the
previous sector and shifted to the right to such a distance to
cover the maximum part of the strip (Fig. 1$a$). These two sectors
will be called a \emph{pair} of sectors. Using sector pairs it is
possible to construct a regular cover with a tile in the form of a
rectangle (in Fig. 1$a$ is a rectangle $BAFE$) whose height
coincides with the width of the strip and is equal to 1 and the
width is $a$.\\

\textbf{Lemma 1.} The density of the cover S1 is determined as
\begin{equation}\label{e1}
    D_1(R,\alpha)=\frac{R^2\alpha}{\sqrt{R^2-(1-\frac{R}{2}\tan\alpha)^2}-\frac{1}{\tan\alpha}+\frac{R}{2}}
\end{equation}
and
$$
\min\limits_{0<\alpha\leq\pi/3;R\sin\alpha\geq
1}D_1(R,\alpha)=\frac{2\pi}{3\sqrt{3}}\approx 1.2092,
$$
when $\alpha=\pi/3=60^\circ$ and $R=2/\sqrt{3}\approx 1.1547$\\

\begin{figure}
\centering
\includegraphics[bb= 25 0 580 500, clip, scale=0.43]{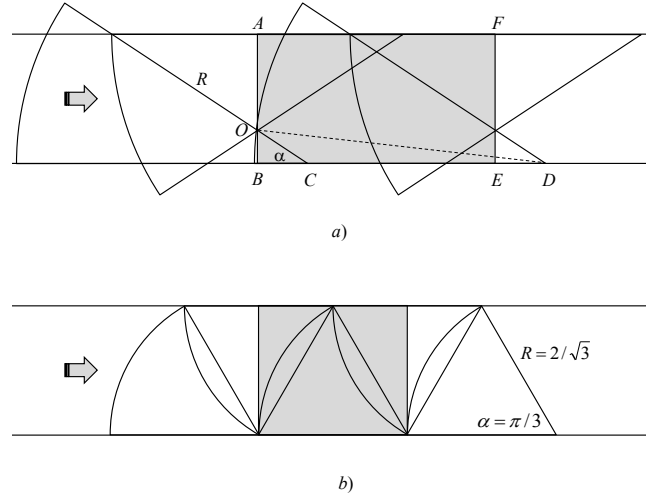}
\caption{$a$) Coverage model S1; $b$) Optimal cover S1.}
\label{fig:fig1}
\end{figure}

\textbf{Proof.} To illustrate the proof, we refer to Fig. 1$a$.
It's obvious that
$$
|BC|=|ED|,\ |AO|=\frac{R}{2}\tan\alpha
$$
and
$$
|OB|=1-|AO|=1-\frac{R}{2}\tan\alpha.
$$
It follows that
$$
|BC|=\frac{|OB|}{\tan\alpha}=\frac{1-\frac{R}{2}\tan\alpha}{\tan\alpha}=\frac{1}{\tan\alpha}-\frac{R}{2}
$$
and
$$
|BD|=\sqrt{R^2-|OB|^2}=\sqrt{R^2-\left(1-\frac{R}{2}\tan\alpha\right)^2}.
$$
Then the width of the tile is
$$
a=|BD|-|ED|=\sqrt{R^2-\left(1-\frac{R}{2}\tan\alpha\right)^2}-\frac{1}{\tan\alpha}+\frac{R}{2}.
$$
Therefore, the density of the cover is
$$
D_1(R,\alpha)=\frac{R^2\alpha}{a}
$$
and after substituting $a$, we obtain (\ref{e1}).

Further it is convenient to represent $R$ in the form
$$
R=1/\sin\alpha+\delta,\ \delta\geq 0.
$$
After substituting $R$
into (\ref{e1}), we obtain the density function
$D_1(\alpha,\delta)$, which is growing by $\delta$. Then the
optimal value is $\delta=0$ and, therefore, $R=1/\sin\alpha$, and
we get the density function
$$
D_1(\alpha)=\frac{\left(\frac{1}{\sin\alpha}\right)^2\alpha}{\sqrt{\left(\frac{1}
{\sin\alpha}\right)^2-\left(1-\frac{1}{2\cos\alpha}\right)^2}-\frac{\cos\alpha}{\sin\alpha}+\frac{1}{2\sin\alpha}}.
$$
The minimum of the density is
$$
\min\limits_{\alpha>0}D_1(\alpha)\approx 1.1767,
$$
when
$$
\alpha=\alpha_1\approx 1.1418 > 65^\circ.
$$
But if $\alpha>\pi/3$, then the model S1 is not a cover.

However, the function $D_1(\alpha)$ decreases, when $\alpha\in
(0,\pi/3]$. Then the optimal value of the angle is $\alpha=\pi/3$,
and
$$
\min\limits_{0<\alpha\leq\pi/3;R\sin\alpha\geq
1}D_1(R,\alpha)=D_1(2/\sqrt{3},\pi/3)=\frac{2\pi}{3\sqrt{3}}.
$$

The proof is over.\\

The optimal cover for the model S1 is shown in Fig. 1$b$.

\subsubsection*{Remark} Since function $D_1(\alpha)$ decreases,
then if $\alpha\in(0,\alpha_1]$, $\alpha_1<\pi/3$, then the
minimum density is reached when $\alpha=\alpha_1$ and
$R=1/\sin\alpha_1$, and it equals
$$
\frac{2\alpha_1}{2\sin^2\alpha_1\sqrt{\frac{1}{\sin^2\alpha_1}-
\left(1-\frac{1}{2\cos\alpha_1}\right)^2}-\sin2\alpha_1+\sin\alpha_1}.
$$
\\

\section{Coverage Model S2}
In model S2, because the angle $\alpha\geq\pi/3$ is greater than
in model S1, two successive sectors with a side at one boundary of
the strip do not intersect (Fig. 2). A pair of sectors symmetrical
with respect to the center line covers a tile ($EACF$ in Fig. 2)
that has a height of 1 and a width of $a=|AC|$.\\

\textbf{Lemma 2.} The density of the cover S2 is determined by
function
\begin{equation}\label{e2}
    D_2(R,\alpha)=\frac{R^2\alpha}{2\sqrt{R^2-1}}
\end{equation}
and its minimum
$$
\min\limits_{\pi/3\leq\alpha\leq\pi/2;R\sin\alpha\geq
1}D_2(R,\alpha)=\pi/3\approx 1.0472,
$$
when $\alpha=\pi/3$ and $R=\sqrt{2}$.\\

\begin{figure}
\centering
\includegraphics[bb= 0 0 530 350, clip, scale=0.45]{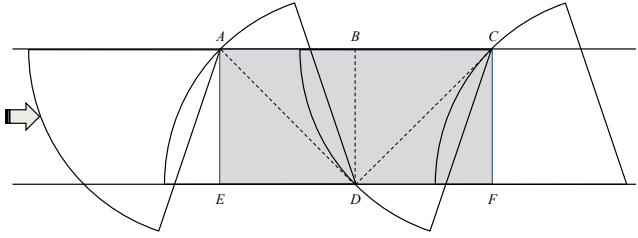}
\caption{Coverage model S2.} \label{fig:fig2}
\end{figure}

\textbf{Proof.} Refer to Fig. 2. It's obvious that
$$
|AD|=|CD|=R
$$
and
$$
|AB|=|BC|=a/2=\sqrt{R^2-1}.
$$
Then the density function
$$
D_2(R,\alpha)=\frac{R^2\alpha}{a}=\frac{R^2\alpha}{2\sqrt{R^2-1}}
$$
decreases with respect to the angle $\alpha\in [\pi/3,\pi/2]$.
Then the optimal value of the angle is $\alpha=\pi/3$, and we get
the density function
$$
D_2(R)=\frac{R^2\pi}{6\sqrt{R^2-1}}.
$$
This is a convex function that reaches the minimum $\pi/3$ when
$R=\sqrt{2}$.

The proof is over.

\section{Coverage Model S3}
The last coverage model, which we consider in this paper, is model
S3, which differs from the previous models in that the sensors are
located on the same boundary of the strip (Fig. 3$a$).\\

\textbf{Lemma 3.} The density of the cover S3 is determined as a
function
\begin{equation}\label{e2}
    D_3(R,\alpha)=\frac{R^2\alpha}{2\left(\sqrt{R^2-1}-\frac{1}{\tan\alpha}\right)}
\end{equation}
and
$$
\min\limits_{0<\alpha\leq\pi/2;R\sin\alpha\geq
1}D_3(R,\alpha)=\pi/2\approx 1.5708,
$$
when $\alpha=\pi/2$ and $R=\sqrt{2}$.\\

\begin{figure}
\centering
\includegraphics[bb= 0 0 580 600, clip, scale=0.5]{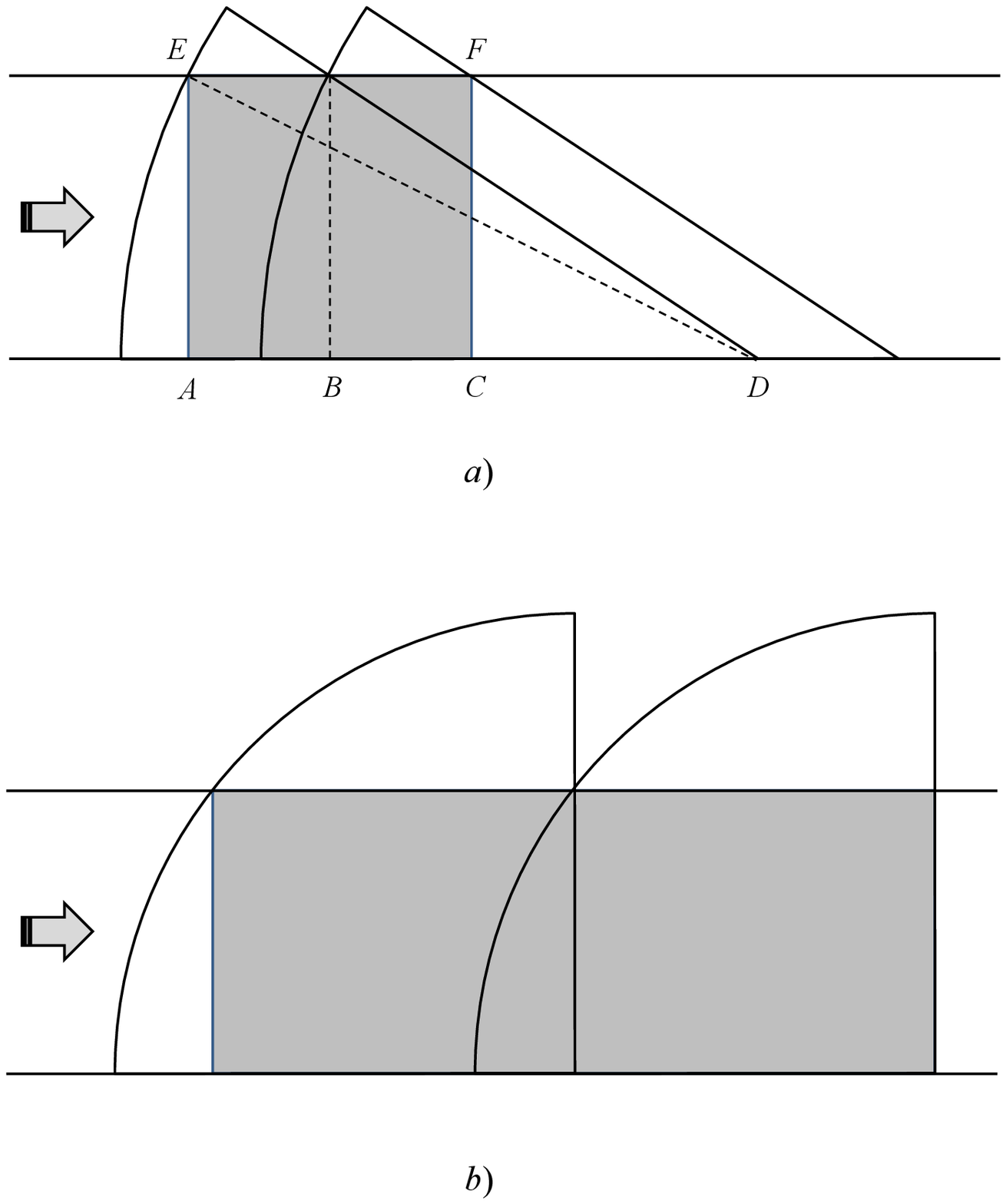}
\caption{$a$) Coverage model S3; $b$) Optimal cover S3.}
\label{fig:fig3}
\end{figure}

\textbf{Proof.} Refer to Fig. 3$a$. The tile is $AEFC$ with height
equals 1 and width $a=|AC|$. We have that
$$
|AB|=|BC|,\ |AD|=\sqrt{R^2-1}
$$
and
$$
|BD|=\frac{1}{\tan\alpha}.
$$
Then
$$
a=|AB|+|BC|=2(|AD|-|BD|)=
$$
$$
2\left(\sqrt{R^2-1}-\frac{1}{\tan\alpha}\right),
$$
and the density function
$$
D_3(R,\alpha)=\frac{R^2\alpha}{a}=\frac{R^2\alpha}{2\left(\sqrt{R^2-1}-\frac{1}{\tan\alpha}\right)}.
$$
Minimum of this function
$$
\min\limits_{0<\alpha\leq\pi/2;R\sin\alpha\geq
1}D_3(R,\alpha)=\pi/2,
$$
when $R=\sqrt{2}$ and $\alpha=\pi/2$.

The proof is over.\\

The optimal cover for the model S3 is shown in Fig. 3$b$.

\section{Conclusion}
Comparing the proposed coverage models of the strip with identical
sectors $(R,\alpha)$, $\alpha\in(0,\pi/2]$, $R\geq 1/\sin\alpha$,
we can draw the following conclusions.
\begin{itemize}
    \item If the angle $\alpha$ can take any values from the domain
    $(0,\pi/2]$, then the best coverage model is model S2, which gives a coverage of the minimum density
    $$
    \min\limits_{\alpha\in(0,\pi/2];R\geq 1/\sin\alpha}D_2(R,\alpha)=\frac{\pi}{3}\approx 1.0472,
    $$
    when $\alpha=\pi/3=60^\circ$ and $R=\sqrt{2}\approx
    1.4142$.
    \item If the angle $\alpha\in(0,\alpha_1]$, $\alpha_1<\pi/3$
    and $R\geq 1/\sin\alpha$,
    then model S2 is inadmissible and model S1 should be used
    with $\alpha=\alpha_1$ and $R=1/\sin\alpha$. In this case, the minimum coverage density
    $$
    \min\limits_{\alpha\in(0,\alpha_1];R\geq 1/\sin\alpha}D_1(R,\alpha)
    $$
    cannot be less than
    $$
    \frac{2\pi}{3\sqrt{3}}\approx 1.2092.
    $$
    \item If the sensors can be placed only on one side of the strip, then model S3 should be
    used. In this case, the minimum coverage density
    $$
    \min\limits_{\alpha\in(0,\pi/2];R\geq 1/\sin\alpha}D_3(R,\alpha)=\pi/2\approx 1.5708,
    $$
    when $\alpha=\pi/2$ and $R=\sqrt{2}$.
\end{itemize}

Since it is necessary to observe the strip from a certain side (in
our case, from right to left), sectors with angles greater than
$\pi/2$ can be located inside the strip and should be oriented so
that the covered points are to the left of the sensor. It can be
shown that in this case it is more efficient to locate the sensors
on the midline of the strip by orienting the sectors symmetrically
relative to the midline (Fig. 4$a$). But such a cover can be
reduced to the cover with sectors with an acute angle, which cover
half the strip (Fig. 4$b$). But this cover is a cover S3 which
covers a half of a strip. The density does not change.

In the strip coverings by identical sectors, which have been
discussed earlier in \cite{ErzinShab15,Erzin16}, and in which it
is not important which sector covers this or that point of the
strip, it was not possible to find the optimum values of the
sensor parameters for any cover model. Consequently, it was not
possible to determine the best coverage model. For the models
considered in this paper, it is possible to find the optimum
values of the sensor parameters. Moreover, in the case when the
parameters $\alpha\in(0,\pi/2]$, $R\geq 1/\sin\alpha$ of the
sensor $(R,\alpha)$ can take arbitrary feasible values, the model
S2 is preferable.

Let's compare the effectiveness of the covers. Since the sensing
energy consumption is proportional to the sensor's coverage area,
cover S1 consumes $1.2092-1.0472=0.162$ extra energy per unit time
than the cover S2. Thus, the WSN in which the S2 coverage model is
used has a lifetime of 16\% more than the sensor network in which
the S1 cover is used, and of 52\% more than the sensor network in
which the S3 cover is used.

\begin{figure}
\centering
\includegraphics[bb= 0 0 580 350, clip, scale=0.5]{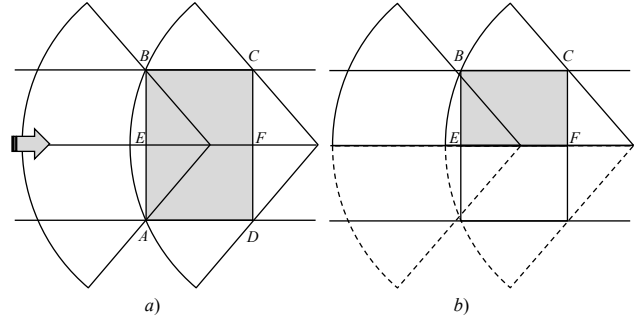}
\caption{$a$) Covering with sectors with an obtuse angle; $b$)
Covering with sectors with an acute angle.} \label{fig:fig4}
\end{figure}

In the case when it is required to monitor the strip in opposite
directions (for example, when people walk along the corridor in
both directions), it is sufficient to add a similar symmetrical
cover to the constructed one, in which the devices are directed in
the opposite directions. Moreover, to save money, the
corresponding devices that are directed in different directions
need to be mounted on a common site.

\subsubsection*{Acknowledgments} This research is supported in part by the Russian
Foundation for Basic Research (grant No. 16-07-00552), the
Ministry of Education and Science of the Republic Kazakhstan
(project No. 0115PK00550) and by Russian Ministry of Science and
Education under the 5-100 Excellence Programme.

For the numerical calculations we used the Maple 17.02 package
licensed to the Novosibirsk State University (serial No.
S2AJ447HV7HAJY5V).

\end{document}